\documentclass{optica-article}
\usepackage[table]{xcolor}

\journal{opticajournal} 

\articletype{Research Article}

\usepackage{lineno}
 
\begin{document}

\title{Measuring the group velocity dispersion in near resonant hot atomic vapors}
\author{Alix Merolle,\authormark{1} Quentin Glorieux\authormark{1,2,*}} 

\address{\authormark{1}Laboratoire Kastler Brossel, Sorbonne Universit\'e, CNRS,
ENS-PSL Research University, Coll\`ege de France, 4 Place Jussieu, 75005 Paris, France\\
\authormark{2}Institut Universitaire de France (IUF)}

\email{\authormark{*}quentin.glorieux@sorbonne-universite.fr}


\begin{abstract*} 
Group velocity dispersion (GVD) in near-resonant hot atomic vapors is difficult to measure with standard pulse broadening or interferometric techniques, as absorption, pulse distortion and nonlinearities strongly affect the probe and reduce the signal-to-noise ratio. 
We introduce a simpler method using a continuous-wave laser with weak phase modulation and a slow photodetector, directly inspired by Bragg-like spectroscopy in fluids of light.
During propagation, the red and blue-detuned sidebands accumulate different dispersive phase shifts, leading to oscillations in the transmitted modulation contrast as the modulation frequency is scanned.
Vanishing contrast at well-defined frequencies directly yields the GVD.
We apply this technique to hot rubidium vapors and observe the strong frequency dependence of the GVD across a broad detuning range of the D2 line at different temperatures.
\end{abstract*}
\section{Introduction and presentation of the technique}
Measuring the dispersion in resonant media is central to slow- and fast-light experiments \cite{boyd2009slow}, nonlinear propagation \cite{glorieux2011quantum} and multiple-wave mixing \cite{glorieux2012generation}. 
In this regime, the GVD is large, the susceptibility is non-linear and the absorption is non-negligible, therefore common techniques such as phase shift measurement \cite{Costa82}, time-of-flight~\cite{bigelow2003observation}, white-light interferometry~\cite{sainz1990refractometry,diddams1996dispersion}, amplitude modulation  \cite{christensen1993simple,devaux1993simple} and pulse broadening~\cite{cohen2003comparison} are complicated by the presence of a nonlinear phase, losses, and spectral filtering. 
We address this by proposing a novel measurement technique using a weakly phase-modulated continuous-wave laser and reading-out the phase evolution of two weak sidebands that co-propagate with the carrier.

It is well known that phase modulation converts into intensity modulation during propagation in dispersive media \cite{chraplyvy1986phase}.
This idea has been exploited for GVD measurements either by converting pulses trains into a CW laser \cite{yamamoto2010group}, or by measuring the AC/DC ratio in a phase-modulated laser \cite{Yamabayashi96}.
In this work, we follow a similar idea and  expand these techniques in order to work in the regime of resonant media by using a method directly inspired by the paraxial fluids of light framework \cite{GLORIEUX2025157}.

Paraxial fluids of light provide an analogy between the transverse dynamics of a laser beam propagating in a nonlinear medium and the collective behavior of a quantum gas. 
This correspondence links nonlinear optics to ultracold atomic physics and benefits from the extensive theoretical and experimental literature of both fields. 
In particular, a standard measurement is the retrieval of the Bogoliubov dispersion of weak excitations \cite{fontaine2018observation}, which is commonly accessed using Bragg spectroscopy \cite{PhysRevLett.82.4569}.
An optical analogue of Bragg spectroscopy has recently been developed for paraxial fluids of light~\cite{PhysRevLett.121.183604}, offering an intuitive way to understand the technique presented in this work.
Let us consider the evolution along the propagation axis $z$ of a homogeneous laser beam (identified as the mean field) on which we imprint two counter-propagating sinusoidal perturbations, for instance along the transverse direction $x$. 
In the linear regime, the Bogoliubov dispersion predicts that such perturbations follow a parabolic (massive) dispersion and therefore propagate along $x$ (with $z$ playing the role of effective time) at a phase velocity:
\[
v_{\phi} = \pm \alpha k_x ,
\]
where $k_x$ is the perturbation wavevector and $\alpha$ is a proportionality constant (equal to the inverse laser wavevector $1/(2k_0)$ in the present case).
At a fixed propagation distance $L$, the resulting interference contrast depends on $k_x$ and on the initial perturbations relative phase.
For example, a pure phase modulation corresponds to a $\pi/2$ phase shift between the 2 sidebands and therefore leading to zero contrast at $z=0$.
After setting the initial conditions, one can thus determine the value of $\alpha$ by scanning $k_x$ and  looking at the contrast at a distance $L$.

In this work, we propose an analogue of this Bragg spectroscopy in the temporal domain. 
For a non-monochromatic laser and ignoring the transverse evolution, we write the propagation equation for the electric field slowly-varying envelope $\mathcal{E}$ in presence of dispersion as~\cite{Larre2016}:
\begin{equation}
\label{eq1}
i \frac{\partial \mathcal{E}}{\partial z} = \frac{D_0}{2} \frac{\partial^2 \mathcal{E}}{\partial t^2} - \frac{i}{v_g} \frac{\partial \mathcal{E}}{\partial t},
\end{equation}
where \(D_0\) is the GVD and $v_g$ the group velocity.
It is then possible to remove the rigid drift term of Eq.~\eqref{eq1}: $\frac{i}{v_g} \frac{\partial \mathcal{E}}{\partial t}$, by going into the reference frame co-moving at $v_g$ with the transform $t'=t-z/v_g$ \cite{glorieux2025paraxialfluidslight}.
We then obtain the canonical equation: 
\begin{equation}
\label{eq2}
i \frac{\partial \mathcal{E}}{\partial z} = \frac{D_0}{2} \frac{\partial^2 \mathcal{E}}{\partial t'^2}.
\end{equation}
It is important to note that the roles of $z$ and $t$ are inverted with respect to a typical Schrödinger equation, $z$ being analogous to a \textit{temporal} coordinate and $t'$ being the effective \textit{spatial} dimension, and thus we will use the corresponding vocabulary.
Therefore, by imprinting two counter-propagating excitations on a homogeneous background (in the co-moving frame), i.e. two small perturbations at $\omega_0 \pm \delta \omega$, with $\omega_0$ being the laser central frequency, we reproduce a similar Bragg configuration but in the time domain as shown in Fig.~\ref{fig:setup}c. 
Interestingly, here the coefficient $\alpha$ that links the phase velocity of the perturbation $v_{\phi}^t$ to the \textit{wavevector} of the perturbation $\delta \omega$ (along $t'$) is directly $D_0$ such that: $v_{\phi}^t=\pm  D_0 \delta \omega/2$.

The GVD can then be estimated directly by measuring the contrast after propagation for a length $L$, while scanning the perturbation wavevector $\delta\omega$. 
The full reconstruction of the $v_{\phi}^t(\delta\omega)$ is possible, but is actually not required since its a linear relation. 
For example, the value of $D_0$ can be retrieved from the first cancellation of contrast during a ramp of $\delta\omega$ from 0. 
In particular, if we impose an initial phase-only modulation (with zero contrast), the counter-propagating perturbations will be destructively interfering again after they both move of half-wavelength (along $t'$) in opposite directions.
We obtain directly the relation between the \textit{wavevector} $\delta\omega_{\text{min}}$ corresponding to the first minimum of contrast and $D_0$:
\begin{equation}
\label{D0eq3}
    D_0=\frac{2\pi}{\delta\omega_{\text{min}}^2L}.
\end{equation}

\section{Experimental description}
\begin{figure*}[ht!]
    \centering
    \includegraphics[width=0.95\textwidth]{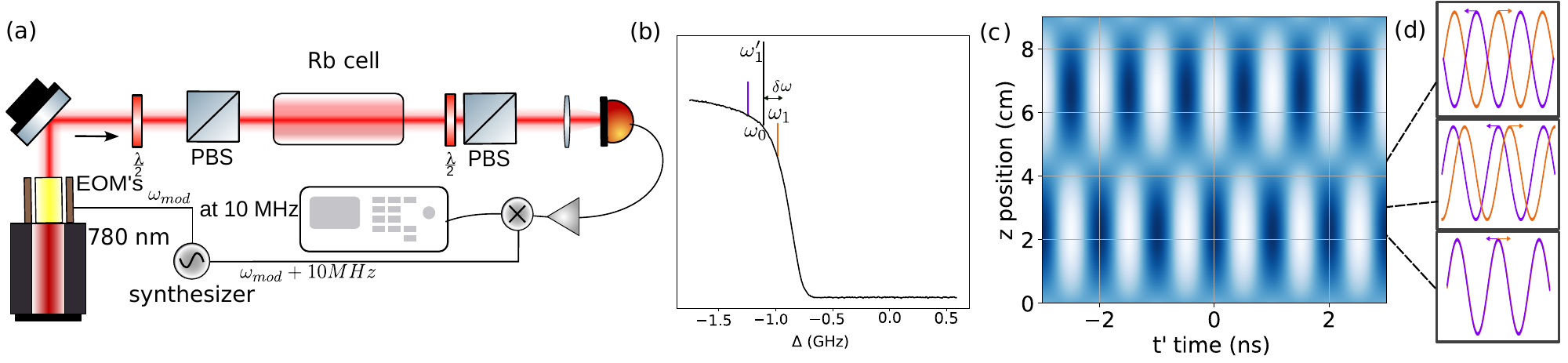}
    \caption{(a) Experimental setup (b) Probe spectrum showing carrier and modulation sidebands in a rubidium Doppler-broaden absorption line. (c) Simulated electric field intensity for medium of 9 cm with a modulation of 4 GHz. The time is represented in the comoving frame $t'$. (d) Interference inside the cell with examples of destructive/constructive phases.}
    \label{fig:setup}
\end{figure*}

In order to realize a temporal Bragg spectroscopy~\cite{PhysRevLett.82.4569}, we use a 780~nm diode laser with its frequency actively stabilized using a high-precision wavemeter, tuned near the D$_2$ transition of $^{87}$Rb in a warm rubidium vapor cell containing a natural mixture of $28\%$ $^{87}$Rb and $72\%$ $^{85}$Rb.  
The cell, sketched in Fig.~\ref{fig:setup}(a), is heated to a controlled temperature between 95~$^\circ$C and 120~$^\circ$C, and for the experiment we employ a vapor cell of length 9~cm.

An electro-optic modulator (EOM), driven by a dual-channel synthesizer, generates two weak sidebands at frequencies $\pm\delta\omega$ around the carrier. 
The carrier is set at a detuning $\Delta$ from  $^{87}$Rb F=2 to F' transition.
The sidebands $\delta\omega$ will then be swept to change the  perturbation \textit{wavevector} (it is a \textit{wavevector} along the $t'$ axis seen as a spatial coordinate) and are typically scanned from $2\pi\times5$~MHz up to $2\pi\times4$~GHz. 
The resulting probe spectrum, including the carrier and its two modulation sidebands, is illustrated in Fig.~\ref{fig:setup}(b) on the side of the Doppler-broaden absorption line of a warm rubidium vapor cell.  
The input modulation depth is kept small, ensuring that only the first-order sidebands at $\pm\delta\omega$ carry appreciable power.
Under these conditions the laser experiences the linear optical response of the vapor: the generated sidebands propagate as weak perturbations whose evolution is governed solely by the frequency-dependent refractive index.
Higher-order harmonics are strongly suppressed, and any nonlinear modification of the refractive index is negligible, so the measured contrast directly reveals the intrinsic dispersion of the medium.
As the sidebands counter-propagate (in the co-moving frame) through the dispersive medium, the initial phase modulation is converted into an effective intensity modulation~\cite{articlemodul}.
Figure~\ref{fig:setup}(c) shows a numerical simulation of the resulting temporal density pattern for a 9~cm medium and a 4~GHz modulation frequency, illustrating the oscillations (along $z$) generated by the counter-propagating excitations.  
As already mentioned, the propagation coordinate $z$ maps to an effective time, with the physical time (in the comoving frame) $t'$ acting as a spatial-like coordinate.  

To ensure operation in the linear regime, all measurements are performed with an incident power of 15~mW and a beam waist of 1.3~mm.
At this level, the intensity inside the vapor cell remains well below the saturation intensity of the resonant rubidium D$_2$ transition, although nearby detuned transitions may be slightly affected, so that absorption and dispersion can still be described within the linear susceptibility of the medium.

This prevents additional nonlinear effects, such as power broadening or Kerr-type phase shifts~\cite{articleboyd}, from significantly distorting the measured contrast and ensures that the observed oscillations originate predominantly from the dispersive properties of the atomic vapor.

The output signal is collected after the medium on a photodiode with bandwidth of 5~GHz.
The detected photocurrent is mixed with a reference signal phase-locked with the EOM driver and frequency-shifted by $+10$~MHz as shown in Fig.~\ref{fig:setup}(a). 
After the mixing stage, the signal is demodulated to retrieve the output spectrum using a spectrum analyzer in zero span mode at an analysis frequency of 10~MHz.
This heterodyne detection scheme isolates the beat note corresponding to the modulation sidebands while rejecting low-frequency technical noise.  
The sidebands frequency ramp triggers the spectrum analyzer acquisition and each time-trace is averaged over 100 acquisitions to further reduce the noise.  
The interference between the counter-propagating sidebands inside the cell, together with examples of constructive and destructive phases, is illustrated in Fig.~\ref{fig:setup}(c-d), highlighting the mechanism responsible for the observed contrast oscillations.

\section{Contrast measurements, GVD and parabolic dispersion retrieval}
\begin{figure}[ht!]
    \centering
    \includegraphics[width=0.99\linewidth]{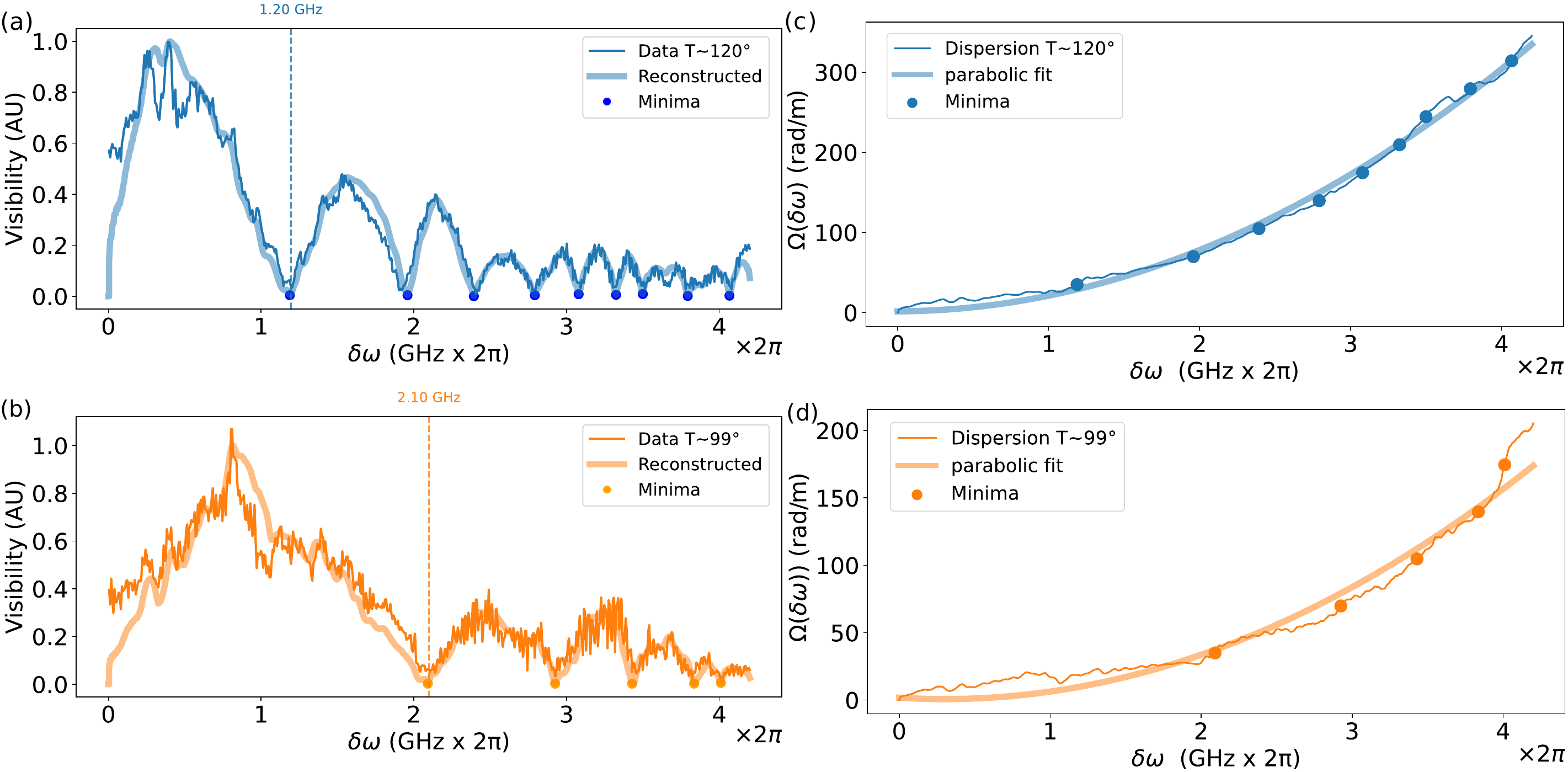}
    \caption{(a) Measured contrast at a cell temperature of \(120^{\circ}\mathrm{C}\) for a detuning \(\Delta=-3.70\;\mathrm{GHz}\) and an input power of 15 mW.  
    Thin blue line: directly measured spectrum; thick blue line: reconstructed contrast after Hilbert-phase analysis. Blue dots: experimental minima 
    (b) Same measurement as in Fig.(a) but for a cell temperature of \(99^{\circ}\mathrm{C}\) (orange curves).  
    (c) Dispersion relation \(\omega(k_t)\) extracted from the minima of the contrast for the \(120^{\circ}\mathrm{C}\) data.  
    Blue dots: experimental points \(\omega(k_t)=p\pi/L\) for successive minima; thin blue line: continuous dispersion retrieved from the reconstructed contrast; thick blue line: parabolic fit.  
    (d) Same analysis as in Fig.(c) for the \(99^{\circ}\mathrm{C}\) data (orange).  
    These panels illustrate how both the contrast spectra and the derived dispersions depend on the vapor temperature.}
    \label{fig:compil}
\end{figure}
From the demodulated signal, we extract the  contrast as a function of the sidebands frequencies $\delta\omega$.
The results are presented in Fig.~\ref{fig:compil}(a) for a cell temperature of \(120^{\circ}\mathrm{C}\) and in Fig.~\ref{fig:compil}(b) for \(99^{\circ}\mathrm{C}\), both recorded at a detuning \(\Delta = -3.7~\mathrm{GHz}\) from the $^{87}$Rb $D_2$ resonance frequency.  
In the high–temperature case of Fig.~\ref{fig:compil}(a), the first minimum occurs at \(\delta\omega_{\text{min}} =2\pi \times  1195~\mathrm{MHz}\), whereas for the cooler cell of Fig.~\ref{fig:compil}(b) the first minimum is shifted to \(\delta\omega_{\text{min}} = 2\pi \times 2099~\mathrm{MHz}\).  
Beyond the first minimum, the positions of the higher-order minima are similarly shifted, consistently reflecting the temperature-dependent increase of the atomic density, which strengthens the linear dispersive response of the medium. 
Following Eq.~\eqref{D0eq3}, this clear shift of the minima directly evidences the stronger GVD at higher atomic density. To emphasize the simplicity and robustness of the measurement, the contrast traces are shown as raw signals directly recorded on the spectrum analyzer.
Building on this observation, one can go beyond the first minimum and exploit the full set of successive minima.
To follow the principle of Bragg spectroscopy, each minimum provides a discrete sampling point of the excitation dispersion. Using these points, we reconstruct the Bogoliubov dispersion relation in Fig.~\ref{fig:compil}(c) (for $120^{\circ}\mathrm{C}$) and Fig.\ref{fig:compil}(d) (for $99^{\circ}\mathrm{C}$).
These minima occur when:
\begin{equation}
\label{p}
\Omega(\delta\omega)\ = p \frac{\pi}{L},
\end{equation}
where \(p\) is an integer and \(L = 9~\mathrm{cm}\) is the cell length and  $\delta\omega$ are the sideband frequencies of the successive contrast minima.  
Knowing the $\delta\omega$ positions of the successive minima therefore gives direct access to the dispersion relation \(\Omega(\delta\omega)\).  
While the minima yield discrete points, the full reconstructed contrast, obtained by smoothing the raw data, normalizing by the envelope, applying a sign-correction, and performing a Hilbert transform, provides the entire dispersion curve continuously for direct comparison with theory.
Figures~\ref{fig:compil}(c-d) show the experimental points (dots), the continuous dispersion retrieved from the reconstructed contrast (thin line), and a parabolic fit (thick line) for the $120^{\circ}\mathrm{C}$ and $99^{\circ}\mathrm{C}$ cells, respectively.  
As expected, the dispersion follows a quadratic behavior, corresponding to the particle–like branch of the Bogoliubov spectrum in the absence of interactions.  
In this representation, the photons behave as non-interacting particles with an effective mass determined by the linear GVD of the medium.  
The curvature of the fitted parabola directly quantifies the GVD, highlighting the temperature-dependent modification of the effective photon mass due to the change in atomic density.  
This approach provides a clear and intuitive mapping between the measured contrast minima and the underlying dispersive properties of the atomic vapor.
Building on this, the GVD at a given detuning can in principle be estimated directly from the first contrast minimum. As shown in Eq.~\eqref{D0eq3}, extracting the frequency of the first minimum 
$\delta\omega_{\text{min}}$ at a given detuning would provide a 
direct estimate of the GVD parameter \(D_0\).  
However, in practice, small residual nonlinearities, which increase as one approaches resonance, can slightly shift the position of the first minimum, even when operating at low input intensity 
For this reason, instead of relying solely on the first minimum, we determine \(D_0(\Delta)\) by performing a parabolic fit to the dispersion relation 
reconstructed from all accessible contrast minima, as shown in 
Figs.~\ref{fig:compil}(c-d).  
The fitted curvature then provides a more robust estimate of the GVD, effectively averaging over several minima and 
suppressing distortions due to weak nonlinear contributions. 
To visualize how the minima evolve with the detuning, we extract all accessible 
contrast minima \(\delta\omega_{\text{min}}^{(n)}(\Delta)\) for each value of 
\(\Delta\).  
\begin{figure}[ht!]
    \centering
    \includegraphics[width=0.99\linewidth]{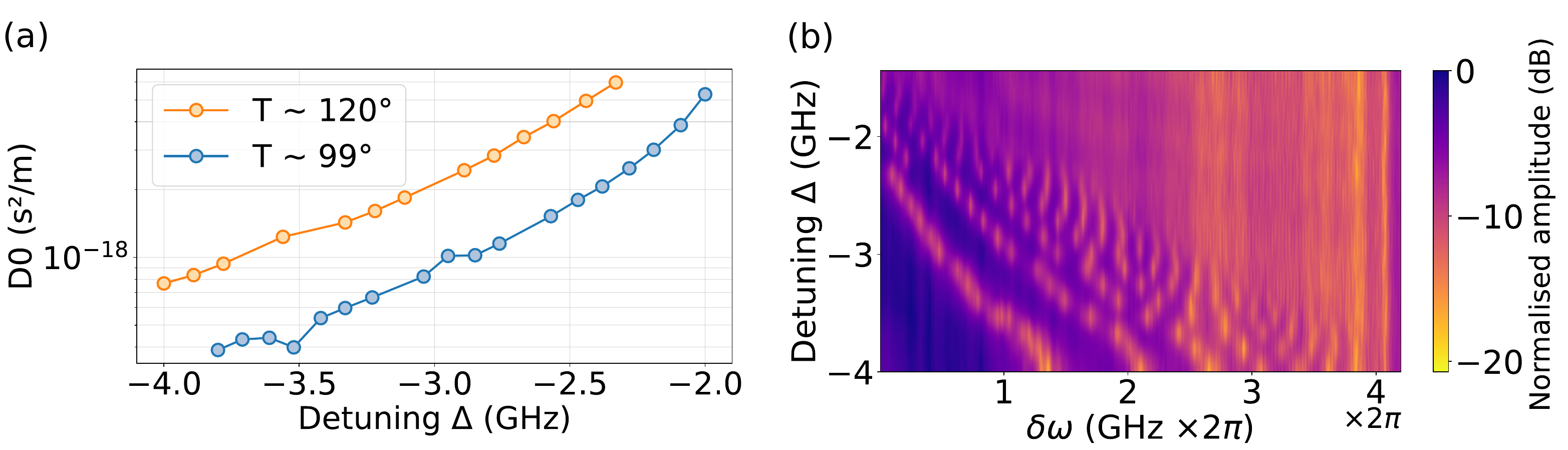}
    \caption{(a) GVD $D_0$  as a function of the detuning $\Delta$. The blue and orange curves represent the results obtained at two  temperatures, respectively \(T = 99^{\circ}\text{C}\) (blue dots) and \(T = 120^{\circ}\text{C}\) (orange dots). The experimental conditions are the same as for the corresponding curves in Fig.~2, except that here the detuning is scanned over the range \(-4\) to \(-2~\text{GHz}\) (x-axe in log scale).
    (b) Positions of the experimentally measured contrast minima $\delta\omega_{\min}^{(n)}$ as a function of the detuning $\Delta$ for several orders $n$, for a cell temperature of $120^{\circ}\mathrm{C}$, illustrating their systematic shift with detuning.}
    \label{fig:D01}
\end{figure}

Their positions are plotted for a cell temperature of $120^{\circ}\mathrm{C}$ in Fig.~\ref{fig:D01}(b) as a function of the 
detuning.  
This representation makes explicit the systematic shift of the minima with 
detuning.
Using these minima, we reconstruct for each detuning the effective dispersion 
relation by performing a parabolic fit to the full set of extracted minima.  
Since the curvature of the quadratic dispersion 
$\Omega(\delta\omega) \simeq \frac{D_{0}}{2}\,\delta\omega^{2}$
is directly set by the second–order GVD coefficient, the fitted parabola provides an immediate and model-independent estimate of $D_{0}$, independent of any specific minimum index.

This procedure, which averages over multiple minima, yields a more robust and 
less noise-sensitive determination of \(D_0\) than methods relying solely on 
the position of the first minimum.
The resulting values of \(D_0(\Delta)\), obtained from this
parabolic-fit procedure, are displayed in Fig.~\ref{fig:D01}(a) in the range \(-4\) to \(-2~\text{GHz}\) for two
temperatures: \(120^{\circ}\mathrm{C}\) (orange dots) and \(99^{\circ}\mathrm{C}\) (blue dots).  
A clear shift between the two curves is observed, reflecting the stronger
dispersive response at higher temperature~\cite{PhysRevLett.116.013601}.  
This temperature dependence is consistent with the expected increase of atomic
density, which enhances the strength of the linear susceptibility and therefore
the magnitude of the GVD near resonance.

\section{Observation of the \(|\Delta|^{3/2}\) scaling law in \(\delta\omega_{\min}\)}
Finally, we explore the dependence of \(\delta\omega_{\min}\) on the detuning. 
To understand this behavior, we consider the effective phase velocities associated with the two modulation sidebands, which can be expressed as:
\begin{equation}
V_{\varphi \pm} = \pm \frac{D_0(\Delta \pm \delta \omega)\,  \delta\omega}{2},
\label{side1}
\end{equation}
where \(D_0(\Delta)\) denotes the GVD coefficient evaluated at the carrier detuning \(\Delta\). Each sideband therefore acquires a slightly different phase velocity depending on whether it is shifted to the red or to the blue of the carrier~\cite{PhysRevLett.83.4277}. 

The contrast minima observed in the experiment occur when the accumulated phase difference between the two sidebands reaches an odd multiple of \(\pi\), resulting in destructive interference after propagation. This condition can be written as  
\begin{equation}
V_{\varphi 1}-V_{\varphi 2} = \frac{2\pi}{L\,\delta\omega_{\min}},
\label{both}
\end{equation}
which provides the phase-matching relation that determines the position of the minima. Recasting Eq.~\eqref{both} yields:  
\begin{equation}
\left[D_0(\Delta+\delta\omega)+D_0(\Delta-\delta\omega)\right]\frac{\delta\omega_{\min}^{2}}{2}
= \frac{2\pi}{L}.
\label{min}
\end{equation}

For small modulation frequencies satisfying \(|\delta\omega|\ll|\Delta|\), we expand \(D_0\) around \(\Delta\) and sum the red- and blue-shifted contributions. The first-order terms \(\pm\delta\omega\,D_0'(\Delta)\) cancel, leaving a second-order correction:
\begin{equation}
D_0(\Delta+\delta\omega)+D_0(\Delta-\delta\omega)
\simeq 2D_0(\Delta) + \delta\omega^{2}D_0''(\Delta).
\label{dev}
\end{equation}

To evaluate Eq.~\eqref{dev}, we use the known frequency dependence of the GVD coefficient in a two-level atomic medium~\cite{glorieux2023hot}. Far from resonance the real part of the susceptibility scales as \(n(\omega)=\mathrm{Re}\,\chi(\omega)\propto 1/\Delta\), which yields
\begin{equation}
D_0 = \frac{d^{2}k}{d\omega^{2}}
\simeq \frac{\omega}{c}\frac{d^{2}n}{d\omega^{2}}
\propto \frac{1}{\Delta^{3}} .
\label{delt3}
\end{equation}

The asymptotic scaling \(D_0 \propto 1/\Delta^{3}\) is valid only in the far-detuned limit \((|\Delta|\gg\gamma)\). Closer to resonance, higher-order corrections to the susceptibility become significant, and the simple algebraic dependence no longer holds.

In the far-detuned regime, the term proportional to \(D_0''(\Delta)\) in Eq.~\eqref{dev} is much smaller than \(2D_0(\Delta)\) and can be neglected. Equation~\eqref{min} therefore simplifies to
\[
2D_0(\Delta)\,\frac{\delta\omega_{\min}^{2}}{2}
= \frac{2\pi}{L}.
\]

Substituting the asymptotic dependence~\eqref{delt3} and solving for \(\delta\omega_{\min}\), we obtain
\begin{equation}
\frac{\delta\omega_{\min}^{2}}{\Delta^{3}}
= \frac{2\pi}{L},
\label{min2}
\end{equation}
from which we directly infer the scaling law
\begin{equation}
\delta\omega_{\min} \propto |\Delta|^{3/2}.
\end{equation}
\begin{figure}[ht!]
    \centering
    \includegraphics[width=0.7\linewidth]{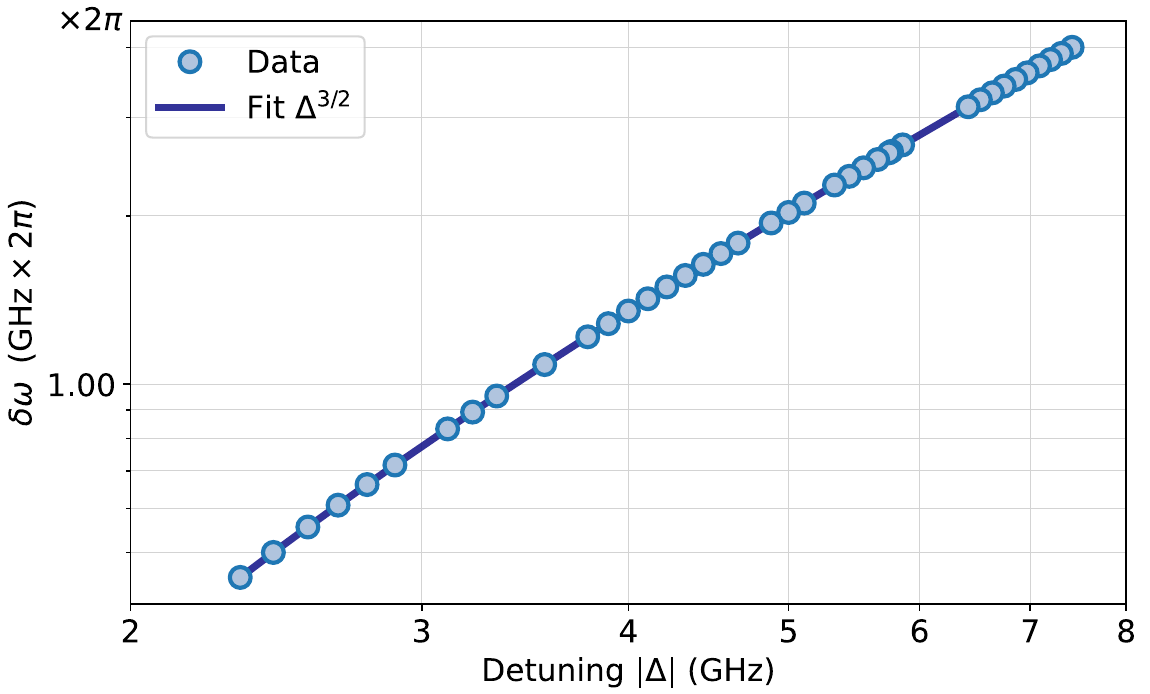}
    \caption{Positions of the first minima $\delta\omega_{\min}$ (blue dots) as a function of detuning at $120^\circ$C. For detunings closer to resonance (>-4 GHz), $\delta\omega_{\min}$ are extracted from the parabolic fit of the reconstructed dispersion curve, while for detunings far from resonance (<-4 GHz) they correspond to the measured contrast minima. The solid blue line shows the power-law fit $\delta\omega_{\min} \propto |\Delta|^{3/2}$ (x- and y-axes in log scale).}
    \label{fig:Ddeltamin}
\end{figure}
To verify this prediction experimentally ,all measurements were performed under the same conditions as in Figs.~\ref{fig:compil}(a) and (c), with a cell temperature of $120^{\circ}\mathrm{C}$ and an input power of 15~mW.
The extracted $\delta\omega_{\min}$ values for each detuning $\Delta$ are shown as blue dots in Fig.~\ref{fig:Ddeltamin}, while the thick blue line represents a power-law fit of the form $\delta\omega_{\min} \propto |\Delta|^{3/2}$. 
In this regime, the experimental points follow the predicted $|\Delta|^{3/2}$ scaling, providing a clear experimental verification of our analytical prediction given by Eq.~\eqref{min2}.
The experimental observation of the \( |\Delta|^{3/2} \) scaling provides a clear validation of our method for extracting \(D_0\) from the reconstructed dispersion curve. 



\section{Conclusion}
In summary, we have introduced a simple and robust technique to measure the group velocity dispersion of hot atomic vapors near resonance using a weakly phase-modulated continuous-wave laser. 
By mapping the oscillations of the transmitted modulation contrast onto a temporal Bragg spectroscopy framework, we obtain direct access to the GVD and its temperature dependence.
This approach circumvents the limitations of conventional pulse-broadening or interferometric methods, which are strongly affected by absorption, pulse distortion and nonlinearities in this regime.
Beyond GVD measurements, the technique opens new possibilities for characterizing the optical response of resonant media and for exploring the dynamics of paraxial fluids of light in strongly dispersive conditions.
\bibliography{main.bib}           
\end{document}